\documentclass[twocolumn, switch]{article}
\usepackage{preprint}

\usepackage{amsmath, amsthm, amssymb, amsfonts}
\usepackage{subfig}
\usepackage{physics}
\usepackage{mathtools}
\usepackage{latexsym}      
\usepackage{graphicx}
\usepackage{upgreek}

\usepackage[numbers,square,sort&compress]{natbib}
\usepackage[utf8]{inputenc}	
\usepackage[T1]{fontenc}	
\usepackage[colorlinks = true,
            linkcolor = purple,
            urlcolor  = blue,
            citecolor = cyan,
            anchorcolor = black]{hyperref}	
\usepackage{booktabs} 		
\usepackage{nicefrac}		
\usepackage{microtype}		
\usepackage{lineno}		    
\usepackage{float}			
\usepackage{stfloats}       

\usepackage{multirow}
\usepackage[table,xcdraw]{xcolor}
\usepackage{lipsum}

\usepackage{newfloat}
\DeclareFloatingEnvironment[name={Supplementary Figure}]{suppfigure}
\usepackage{sidecap}
\sidecaptionvpos{figure}{c}

\usepackage{chemformula} 
\let\ce\ch

\usepackage{titlesec}
\titlespacing\section{0pt}{12pt plus 3pt minus 3pt}{1pt plus 1pt minus 1pt}
\titlespacing\subsection{0pt}{10pt plus 3pt minus 3pt}{1pt plus 1pt minus 1pt}
\titlespacing\subsubsection{0pt}{8pt plus 3pt minus 3pt}{1pt plus 1pt minus 1pt}

\usepackage{tikz,hyperref}
\definecolor{lime}{HTML}{A6CE39}
\DeclareRobustCommand{\orcidicon}{
	\begin{tikzpicture}
	\draw[lime, fill=lime] (0,0)
	circle [radius=0.16]
	node[white] {{\fontfamily{qag}\selectfont \tiny ID}};
	\draw[white, fill=white] (-0.0625,0.095)
	circle [radius=0.007];
	\end{tikzpicture}
	\hspace{-2mm}
}
\foreach \x in {A, ..., Z}{\expandafter\xdef\csname orcid\x\endcsname{\noexpand\href{https://orcid.org/\csname orcidauthor\x\endcsname}
			{\noexpand\orcidicon}}
}

\title{Uncertainty Quantification for Free Energy Calculations by Generalized Hierarchical Bayesian Inference}

\usepackage{authblk}

\author[1]{Martin Skorna\orcidA{}}
\author[1]{Adam Gottfried\orcidB{}}
\author[1]{Zuzana Janackova\orcidC{}}
\author[1]{Katarina Baxova\orcidD{}}
\author[1]{Pavel Jungwirth\orcidE{}}
\author[1]{Brennon L. Shanks*\orcidF{}}

\affil[*]{Correspondence: \ttfamily shanks.brennon@uochb.cas.cz}

\affil[1]{Institute of Organic Chemistry and Biochemistry of the Czech Academy of Sciences, Prague, CZH}

\begin{document}

\twocolumn[ 
  \begin{@twocolumnfalse} 

\maketitle

\begin{abstract}

Free energy calculations are routinely used to study molecular processes inaccessible to unbiased molecular dynamics, but their utility ultimately depends on knowing when and how much their predictions can be trusted. 
Uncertainty estimation is therefore essential for distinguishing genuine physical features of a free energy profile from artifacts arising from limited simulation data or inadequate sampling.
Gaussian processes have emerged as a powerful framework for reconstructing free energy profiles together with predictive uncertainties.
However, existing implementations typically condition on fixed hyperparameters and observation noise, preventing predictive uncertainties from adapting to the information content of the simulation data. 
Here, we develop a generalized hierarchical Gaussian process framework that accounts for these neglected sources of uncertainty. 
Applications to umbrella sampling and extended Lagrangian metadynamics of peptide–lipid membrane interactions demonstrate that the resulting uncertainty estimates track reconstruction errors across a wide range of sampling and data conditions.

\end{abstract}


\vspace{0.35cm}

  \end{@twocolumnfalse} 
] 

\clearpage


\section{Introduction}

Enhanced-sampling free energy calculations are among the most widely used applications of molecular dynamics (MD), providing a powerful framework for understanding atomic-scale interactions and thermodynamic processes in molecular systems.
By projecting the high-dimensional configurational phase space of a molecular system onto a reduced set of coordinates, free energy calculations facilitate the study of molecular processes whose characteristic timescales lie beyond the reach of unbiased MD.
Over the past several decades, the development of enhanced sampling techniques such as umbrella sampling \cite{torrie_monte_1974,kumar_weighted_1992,kastner_bridging_2005}, metadynamics \cite{laio_escaping_2002,laio_metadynamics_2008}, and related methods \cite{shirts_2008,lindahl_accelerated_2014, comer_adaptive_2015, york_modern_2023} has dramatically expanded the range of systems that can be studied computationally.

A fundamental challenge of free energy methods is principled uncertainty quantification (UQ), which aims to estimate both the free energy profile and its uncertainty.
This problem is especially challenging for short simulation trajectories or sparse sampling.
In such low-data regimes, rigorous UQ would enable adaptive sampling strategies that avoid over-sampling of well-characterized regions and under-sampling near barriers or transition states, thus saving valuable computational effort and enabling automated computational workflows.
However, standard UQ approaches rely on assumptions that become questionable in precisely these regimes.
For example, error propagation for variance estimators with sparse umbrella window centers, given a normal distribution approximation for the biased distributions, underestimates the total sampled error \cite{kastner_analysis_2006}.
On the other hand, block averaging and bootstrapping methods require fully equilibrated trajectories and user-specified choice of the number of independent trajectory epochs, which significantly impacts the estimated predictive variance \cite{zhu_convergence_2012}.
While these approaches can be effective under ideal sampling conditions, their uncertainty estimates are less informative when data are limited, precisely where accurate reliable UQ is most important.

To address these limitations, we employ a Gaussian process (GP), which treats free energy reconstruction as a problem of statistical inference.
In a nutshell, the method assumes the free energy is an unknown function of the collective variables, and begins with a broad range of physically plausible functions that satisfy basic assumptions such as continuity and differentiability. 
This initial description, known as the \textit{prior}, represents our knowledge before obtaining simulation data.
MD simulations then provide incomplete information about this unknown free energy, and Bayesian inference quantifies the probability of different functions as new data is incorporated.
The final result is a probability distribution, known as the \textit{posterior}, that identifies the most plausible free energy functions given both the prior assumptions and the simulation data.
GPs have been used for umbrella sampling \cite{stecher_free_2014}, metadynamics \cite{mones_exploration_2016}, mulistate Bennet acceptance ratio \cite{ding_2024}, and the interpolation of quantum potential energy surfaces \cite{dai_interpolation_2020, sugisawa_gaussian_2020}. 
Bayesian optimization for umbrella sampling, which leverages a GP as a guide for biasing window selection, has even been shown to improve sampling efficiency by a factor of 1.6-2.8 across a diverse set of two-dimensional systems \cite{kempkes_bayesian_2026}. 

While existing GP-based free energy estimators have been successfully applied across multiple areas of computational chemistry, rigorous evaluation of their UQ performance remains limited. 
A key unresolved issue is the treatment of hyperparameters, which are GP model parameters that specify properties of the prior. 
Earlier work has either fixed hyperparameters \textit{a priori} based on empirical evidence of low root-mean-square-error (RMSE) between the GP mean and "ground truth" or varied hyperparameters to assess sensitivity \textit{post hoc}. 
As we will show, assuming GP hyperparameters neglects important sources of uncertainty and can prevent the model from adapting as new simulation data are acquired.
Consequently, the resulting uncertainty estimates are often misleading and do not accurately reflect the true level of uncertainty in the reconstruction.

\begin{figure*}[t!]
     \centering
     \includegraphics[width=\linewidth]{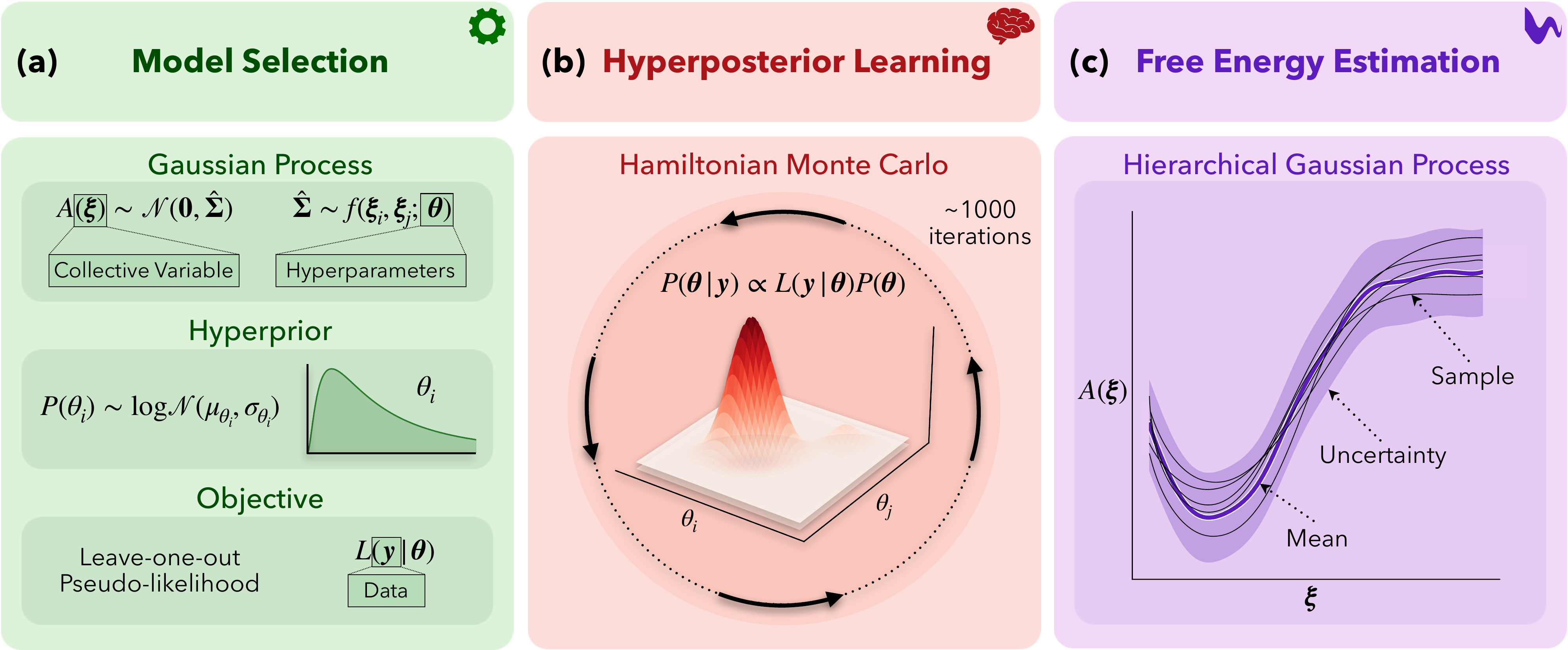}
     \caption{Flowchart summarizing the workflow of the generalized hierarchical Bayesian framework for free energy reconstruction.}
     \label{fig:flowchart}
\end{figure*}

Here, we develop a generalized hierarchical GP framework for free energy reconstruction that provides reliable uncertainty estimates that respond consistently to data quality and availability. 
The method combines a joint function–derivative GP observation model with Hamiltonian Monte Carlo No-U-Turn Sampling (HMC-NUTS) of a generalized hyperposterior to propagate uncertainty in the GP hyperparameters to the free energy. 
Using the free energy profile of interaction of a nonaarginine (\ce{R9}) peptide with a lipid bilayer as a benchmark, an important system for understanding the membrane penetration mechanism of cell-penetrating peptides (CPPs), we find that several commonly employed UQ strategies can produce misleading assessments of convergence, either by exhibiting little sensitivity to additional data or by becoming overconfident as sampling progresses. 
In contrast, the hierarchical Bayesian approach avoids these pathological behaviors and yields uncertainties that systematically evolve with increasing data acquisition. 
Together, these results highlight the importance of hyperparameter uncertainty in free energy UQ and suggest a practical path toward uncertainty-aware and automated free energy workflows.

\section{Theory and Computational Methods}

Enhanced sampling simulations provide samples of the biased probability distribution of a molecular system along a collective variable (CV), $\boldsymbol{\xi}$, which serve as input data $\mathbf{y}$ for GP reconstruction of the free energy profile (Figure~\ref{fig:flowchart}).

Model selection (Figure~\ref{fig:flowchart}a) is the process of specifying a probabilistic model for free energy inference.
This involves choosing the GP prior mean and covariance functions, which depend on unknown parameters called \textit{hyperparameters}.
Priors on these quantities, called \textit{hyperpriors}, constrain nonphysical behavior while allowing the model sufficient flexibility to learn from the MD data.
Because priors are specified for both the free energy profile and the GP hyperparameters, the method is \textit{hierarchical}, consisting of multiple levels of uncertainty that are ultimately propagated to the final prediction.
The final component of model selection is the choice of \textit{likelihood}, which quantifies how well a model explains the observed data.
Here, we employ a leave-one-out (LOO) pseudo-likelihood that favors models with strong predictive performance on held-out observations.

Hyperposterior learning (Figure~\ref{fig:flowchart}b) is the process of sampling the unknown probability distribution given the model and data. 
The unnormalized \textit{hyperposterior}, the posterior distribution over the GP hyperparameters, is given by Bayes' theorem as the product of the hyperpriors and the likelihood.
With the LOO pseudo-likelihood, this object is technically not a proper hyperposterior, but rather a pseudo- or generalized hyperposterior.
The resulting distribution is not available in closed form and is therefore sampled via HMC-NUTS.

Free energy reconstruction (Figure~\ref{fig:flowchart}c) is performed by propagating samples from the generalized hyperposterior through the GP predictive equations, yielding predictive credible intervals for the free energy surface.
The result is commonly summarized by the posterior mean and a $\pm\sigma$ or $\pm 2\sigma$ credible interval, although posterior samples can also be propagated to quantify uncertainty in downstream observables such as free energy barriers.
The resulting free energy posterior therefore reflects uncertainty arising from both the molecular simulation data and the inferred GP hyperparameters.

\subsection{Free Energy Estimation}\label{sec:fe_acq}

The first stage of the workflow is the construction of observations of the free energy profile from enhanced sampling MD simulations. 
Because the underlying theory and practical implementation of free energy methods are well documented elsewhere~\cite{pohorille_good_2010}, we restrict ourselves to a brief discussion of the approaches used in the present work: umbrella sampling and extended Lagrangian metadynamics. 

\paragraph{Umbrella Sampling}

In umbrella sampling, a system is simulated under a set of biased potentials centered at prescribed locations $\boldsymbol{\xi}_w$ in CV space. 
A CV is a function of the microscopic coordinates chosen to describe the molecular process of interest, such as the distance or angle between two species. 
One or more CVs define the reduced coordinate space used for free energy reconstruction.
For a harmonic umbrella window $w$ with force constant $\kappa_w$, the biasing potential is
\begin{equation}
    \psi^w(\boldsymbol{\xi}) = \frac{1}{2}\kappa_w \left( \boldsymbol{\xi} - \boldsymbol{\xi}_w \right)^2.
\end{equation}
The resulting biased simulations provide two complementary sources of information: biased probability histograms and local mean-force estimates.

Umbrella sampling provides empirical observations of the free energy function values, up to an additive constant.
For bin $i$ centered at $\boldsymbol{\xi}_i$, the unbiased free energy is related to the biased histograms by
\begin{equation}
    A(\boldsymbol{\xi}_i) = -\beta^{-1}\log P^w(\boldsymbol{\xi}_i) - \psi^w(\boldsymbol{\xi}_i) + c_w,
\end{equation}
where $P^w(\boldsymbol{\xi}_i)$ is a histogram-based estimate of the biased probability density at bin $i$ from window $w$ and $c_w$ is an unknown window-dependent additive constant arising from the normalization of the biased distribution.
In the weighted histogram analysis method (WHAM), these constants are determined self-consistently across all windows to reconstruct a global unbiased probability distribution~\cite{kumar_weighted_1992}.
WHAM also has a Bayesian interpretation \cite{ferguson_bayeswham_2017}.
In the present framework, we instead treat the window-dependent constants as nuisance GP hyperparameters and marginalize them out analytically, while using the histogram-derived values as local observations of the free energy profile.

The free energy gradient at window $w$ can be estimated from the mean restraining force in the biased simulation
\begin{equation}
    \pdv{A(\boldsymbol{\xi})}{\boldsymbol{\xi}} \bigg|_{\boldsymbol{\xi}_w}
    \approx
    - \big\langle \kappa_w \left( \boldsymbol{\xi} - \boldsymbol{\xi}_w \right) \big\rangle_w,
\end{equation}
where $\big\langle \cdot\big\rangle_w$ is the mean of the biased distribution sampled in window $w$. 
These quantities are treated as free energy derivative observations, in a manner analogous to the umbrella integration method \cite{kastner_bridging_2005}.

\paragraph{Metadynamics}

Metadynamics is an enhanced sampling method in which the system is biased along a set of CVs by depositing history-dependent Gaussian potentials centered on previously visited CV values. 
As the simulation progresses, the accumulated bias progressively destabilizes previously explored free energy minima, encouraging transitions into neighboring regions of CV space. 
Under appropriate conditions, the accumulated bias converges to the negative of the underlying free energy profile (up to a constant).

In extended Lagrangian formulations, auxiliary variables $\boldsymbol{z}$ are introduced and coupled to the CVs through a stiff harmonic restraint \cite{laio_metadynamics_2008},
\begin{equation}
    V_{\mathrm{coupling}}(\boldsymbol{\xi}, \boldsymbol{z}) = \frac{1}{2}\kappa \left( \boldsymbol{\xi} - \boldsymbol{z} \right)^2,
\end{equation}
while the history-dependent bias is applied only in the auxiliary space.
Under sufficiently stiff coupling, $\boldsymbol{z}\approx\boldsymbol{\xi}$, and
\begin{equation}
    \frac{\partial A}{\partial \boldsymbol{z}} \bigg|_{\boldsymbol{z}^*}
    = \left\langle \frac{\partial V_{\mathrm{coupling}}}{\partial \boldsymbol{z}} \right\rangle_{\boldsymbol{z}^*}
    = - \big\langle \kappa \left( \boldsymbol{\xi} - \boldsymbol{z} \right) \big\rangle_{\boldsymbol{z}^*}.
\end{equation}
The physical potential does not depend explicitly on $\boldsymbol{z}$, hence the estimator reduces to the derivative of the coupling term alone. 
This expression provides direct estimates of the mean force at sampled values of $\boldsymbol{z}$, which are treated as noisy derivative observations of the free energy profile.

Histogram observations are obtained by a discrete binning of neighboring $\boldsymbol{z}$ to construct an umbrella biased histogram of the corresponding $\boldsymbol{\xi}$ values.
The binning procedure, which includes choosing the number of bins and the bin width, adds an additional layer of uncertainty, and is discussed in the Supporting Information Section~\ref{si:metad}.

\subsection{Gaussian Process Regression}\label{sec:GPR}

The core of our approach is to model the free energy profile $f(\boldsymbol{\xi})$ over the CV space $\boldsymbol{\xi}\in\mathbb{R}^n$ as a zero mean GP,
\begin{equation}
    f \sim \text{GP}\!\big(\boldsymbol{0}, k(\boldsymbol{\xi}, \boldsymbol{\xi}')\big),
\end{equation}
so that for a set of test inputs $\hat{\boldsymbol{X}}_* = [\boldsymbol{\xi}_1,\dots,\boldsymbol{\xi}_{N_*}]$, the latent function values
\begin{equation}
    \boldsymbol{f}_* = \left[f(\boldsymbol{\xi}_i)\right]_{i=1}^{N_*}
\end{equation}
follow a zero mean multivariate normal distribution with covariance matrix $\hat{\boldsymbol{K}}(\hat{\boldsymbol{X}}_*,\hat{\boldsymbol{X}}_*)$, where $[\hat{\boldsymbol{K}}]_{ij}=k(\boldsymbol{\xi}_i,\boldsymbol{\xi}_j)$ are elements governed by the kernel function.
Given noisy free energy observations $\boldsymbol{y}=[y_1,\dots,y_N]$ at inputs $\hat{\boldsymbol{X}}=[\boldsymbol{\xi}_1,\dots,\boldsymbol{\xi}_N]$, the standard GP posterior mean and covariance is,
\begin{align}
\bar{\boldsymbol{f}}_* \;=\;&
\hat{\boldsymbol{K}}(\hat{\boldsymbol{X}}_*,\hat{\boldsymbol{X}})
\left[\hat{\boldsymbol{K}}(\hat{\boldsymbol{X}},\hat{\boldsymbol{X}})+\sigma_{\mathrm{n}}^2\hat{\mathbf{I}}\right]^{-1}
\boldsymbol{y}, \label{eq:pred_mean} \\
\mathrm{cov}[\boldsymbol{f}_*] \;=\;&
\nonumber \hat{\boldsymbol{K}}(\hat{\boldsymbol{X}}_*,\hat{\boldsymbol{X}}_*)
-
\hat{\boldsymbol{K}}(\hat{\boldsymbol{X}}_*,\hat{\boldsymbol{X}}) \\
& \cdot \left[\hat{\boldsymbol{K}}(\hat{\boldsymbol{X}},\hat{\boldsymbol{X}})+\sigma_{\mathrm{n}}^2\hat{\mathbf{I}}\right]^{-1}
\hat{\boldsymbol{K}}(\hat{\boldsymbol{X}},\hat{\boldsymbol{X}}_*).
\label{eq:pred_cov}
\end{align}

\paragraph{Inference from Free Energy Histograms}

To infer the free energy profile from biased histograms in a manner analogous to WHAM, unknown additive constants associated with individual umbrella windows must be accounted for. 
Following established GP formulations, we represent these offsets through an explicit linear basis in the GP mean. 
For each bin center $\boldsymbol{\xi}_i$, we define a characteristic basis vector
\begin{equation}
    \boldsymbol{h}(\boldsymbol{\xi}_i)
    =
    \left[
    \delta_{\boldsymbol{\xi}_i}(w)
    \;\big|\;
    w=1,\dots,N_w
    \right],
\end{equation}
where
\begin{equation}
    \delta_{\boldsymbol{\xi}_i}(w)=
    \begin{cases}
        1, & \boldsymbol{\xi}_i \in w,\\
        0, & \boldsymbol{\xi}_i \notin w,
    \end{cases}
\end{equation}
indicates whether bin $i$ belongs to window $w$. 
The corresponding weights $\boldsymbol{\beta}$ are the unknown window-dependent offsets. 
Collecting the basis vectors for all training and test points gives the design matrices $\hat{\boldsymbol{H}}$ and $\hat{\boldsymbol{H}}_*$, respectively, with entries $[\hat{\boldsymbol{H}}]_{iw}=1$ if bin $i$ belongs to window $w$ and zero otherwise.

The observation model for histogram-derived free energy values is written as
\begin{equation}
    \boldsymbol{y}
    =
    \hat{\boldsymbol{H}}\boldsymbol{\beta}
    + \boldsymbol{f}(\hat{\boldsymbol{X}})
    + \boldsymbol{\varepsilon},
    \quad
    \boldsymbol{\varepsilon}\sim\mathcal{N}\!\big(\boldsymbol{0},\hat{\boldsymbol{\Sigma}}_{\boldsymbol{y}}\big),
\end{equation}
where $\boldsymbol{\beta}$ represents the unknown window-dependent offsets. 
These nuisance parameters are assigned a Gaussian prior and analytically marginalized, yielding a modified GP posterior.
As shown by O'Hagan~\cite{ohagan_curve_1978}, in the limit of an uninformative prior over $\boldsymbol{\beta}$, the posterior predictive distribution remains Gaussian, with mean and covariance
\begin{equation}\label{eq:g_ohagan}
    \bar{\boldsymbol{g}}_*
    =
    \bar{\boldsymbol{f}}_*
    + \hat{\boldsymbol{R}}^{\mathrm{T}}\bar{\boldsymbol{\beta}},
    \quad
    \mathrm{cov}[\boldsymbol{g}_*]
    =
    \mathrm{cov}[\boldsymbol{f}_*]
    + \hat{\boldsymbol{R}}^{\mathrm{T}}\hat{\boldsymbol{S}}^{-1}\hat{\boldsymbol{R}},
\end{equation}
where
\begin{equation}
\begin{aligned}
\hat{\boldsymbol{R}}
&=
\hat{\boldsymbol{H}}_*^{\mathrm{T}}
-
\hat{\boldsymbol{H}}^{\mathrm{T}}
\left[
\hat{\boldsymbol{K}}(\hat{\boldsymbol{X}},\hat{\boldsymbol{X}})
+\sigma_{\mathrm{f}}^2\hat{\mathbf{I}}
\right]^{-1}
\hat{\boldsymbol{K}}(\hat{\boldsymbol{X}},\hat{\boldsymbol{X}}_*),\\
\bar{\boldsymbol{\beta}}
&=
\left(
\hat{\boldsymbol{H}}^{\mathrm{T}}
\left[
\hat{\boldsymbol{K}}(\hat{\boldsymbol{X}},\hat{\boldsymbol{X}})
+\sigma_{\mathrm{f}}^2\hat{\mathbf{I}}
\right]^{-1}
\hat{\boldsymbol{H}}
\right)^{-1}
\hat{\boldsymbol{H}}^{\mathrm{T}}
\left[
\hat{\boldsymbol{K}}(\hat{\boldsymbol{X}},\hat{\boldsymbol{X}})
+\sigma_{\mathrm{f}}^2\hat{\mathbf{I}}
\right]^{-1}
\boldsymbol{y},\\
\hat{\boldsymbol{S}}
&=
\hat{\boldsymbol{H}}^{\mathrm{T}}
\left[
\hat{\boldsymbol{K}}(\hat{\boldsymbol{X}},\hat{\boldsymbol{X}})
+\sigma_{\mathrm{f}}^2\hat{\mathbf{I}}
\right]^{-1}
\hat{\boldsymbol{H}}.
\end{aligned}
\end{equation}
The additional terms relative to Equations~\eqref{eq:pred_mean} and \eqref{eq:pred_cov} account for the uncertainty associated with the unknown window-dependent offsets.

\paragraph{Inference from Free Energy Gradients}

The same framework can be extended to noisy derivative observations, the main difference being that we no longer need to consider the unknown additive constants arising from the biased probability distributions (they become zero after differentiation). 
Let $\boldsymbol{y}'=[y'_1,\dots,y'_M]$ denote gradient observations associated with input locations $\hat{\boldsymbol{X}}'=[\boldsymbol{\xi}'_1,\dots,\boldsymbol{\xi}'_M]$. 
Since differentiation is linear, derivatives of a GP remain jointly Gaussian. 
Defining the structured derivative operators
\begin{equation}
    [\partial_i \hat{\boldsymbol{K}}(\hat{\boldsymbol{X}},\hat{\boldsymbol{X}})]_{ij}
    =
    \nabla_{\boldsymbol{\xi}_i}k(\boldsymbol{\xi}_i,\boldsymbol{\xi}_j),
\end{equation}
and
\begin{equation}
    [\partial_i\partial_j \hat{\boldsymbol{K}}(\hat{\boldsymbol{X}},\hat{\boldsymbol{X}})]_{ij}
    =
    \nabla_{\boldsymbol{\xi}_i}\nabla_{\boldsymbol{\xi}_j}k(\boldsymbol{\xi}_i,\boldsymbol{\xi}_j),
\end{equation}
the posterior predictive distribution for derivative-informed inference is Gaussian with mean and covariance
\begin{align}
\bar{\boldsymbol{f}}_* \;=\;&
\partial_j\hat{\boldsymbol{K}}(\hat{\boldsymbol{X}}_*,\hat{\boldsymbol{X}}')
\left[
\partial_i\partial_j\hat{\boldsymbol{K}}(\hat{\boldsymbol{X}}',\hat{\boldsymbol{X}}')
+ \sigma_{\mathrm{d}}^2\hat{\mathbf{I}}
\right]^{-1}
\boldsymbol{y}', \label{eq:f'_mean} \\
\mathrm{cov}[\boldsymbol{f}_*] \;=\;&
\nonumber \hat{\boldsymbol{K}}(\hat{\boldsymbol{X}}_*,\hat{\boldsymbol{X}}_*)
-
\partial_j\hat{\boldsymbol{K}}(\hat{\boldsymbol{X}}_*,\hat{\boldsymbol{X}}') \\
& \cdot \left[
\partial_i\partial_j\hat{\boldsymbol{K}}(\hat{\boldsymbol{X}}',\hat{\boldsymbol{X}}')
+ \sigma_{\mathrm{d}}^2\hat{\mathbf{I}}
\right]^{-1}
\partial_j\hat{\boldsymbol{K}}(\hat{\boldsymbol{X}}_*,\hat{\boldsymbol{X}}')^{\mathrm{T}}.
\label{eq:f'_cov}
\end{align}

\paragraph{Joint Histogram-Gradient Inference}

Both observation types can be incorporated into a single GP model by concatenating the observation vectors, $\boldsymbol{y}\oplus\boldsymbol{y}'$, and constructing the joint latent covariance matrix
\begin{equation}
    \hat{\boldsymbol{K}}_{\boldsymbol{y}\oplus\boldsymbol{y}'}
    =
    \begin{pmatrix}
        \hat{\boldsymbol{K}}(\hat{\boldsymbol{X}},\hat{\boldsymbol{X}})
        &
        \partial_j\hat{\boldsymbol{K}}(\hat{\boldsymbol{X}},\hat{\boldsymbol{X}}') \\
        \partial_i\hat{\boldsymbol{K}}(\hat{\boldsymbol{X}}',\hat{\boldsymbol{X}})
        &
        \partial_i\partial_j\hat{\boldsymbol{K}}(\hat{\boldsymbol{X}}',\hat{\boldsymbol{X}}')
    \end{pmatrix}.
\end{equation}
The corresponding test--training cross-covariance is
\begin{equation}
    \hat{\boldsymbol{K}}_{*,\,\boldsymbol{y}\oplus\boldsymbol{y}'}
    =
    \begin{bmatrix}
        \hat{\boldsymbol{K}}(\hat{\boldsymbol{X}}_*,\hat{\boldsymbol{X}})
        &
        \partial_j\hat{\boldsymbol{K}}(\hat{\boldsymbol{X}}_*,\hat{\boldsymbol{X}}')
    \end{bmatrix}.
\end{equation}
The joint observation noise covariance is
\begin{equation}
    \hat{\boldsymbol{\Sigma}}_{\boldsymbol{y}\oplus\boldsymbol{y}'}
    =
    \begin{pmatrix}
        \hat{\boldsymbol{\Sigma}}_{\boldsymbol{y}} & \hat{\mathbf{O}} \\
        \hat{\mathbf{O}} & \hat{\boldsymbol{\Sigma}}_{\boldsymbol{y}'}
    \end{pmatrix},
\end{equation}
where $\hat{\boldsymbol{\Sigma}}_{\boldsymbol{y}}$ and $\hat{\boldsymbol{\Sigma}}_{\boldsymbol{y}'}$ denote the covariance matrices associated with the histogram and derivative observations, respectively. In this work, the histogram and derivative observation noise are assumed to be uncorrelated, so the off-diagonal blocks are zero. The corresponding joint observation covariance is therefore
\begin{equation}
    \hat{\boldsymbol{C}}_{\boldsymbol{y}\oplus\boldsymbol{y}'}
    =
    \hat{\boldsymbol{K}}_{\boldsymbol{y}\oplus\boldsymbol{y}'}
    +
    \hat{\boldsymbol{\Sigma}}_{\boldsymbol{y}\oplus\boldsymbol{y}'}.
\end{equation}

The unknown window-dependent offsets apply only to the histogram observations and vanish for derivative observations. The posterior predictive mean and covariance then follow directly from Equations~\eqref{eq:pred_mean} and \eqref{eq:pred_cov} by replacing the observation vector $\boldsymbol{y}$ with $\boldsymbol{y}\oplus\boldsymbol{y}'$, the training covariance $\hat{\boldsymbol{K}}(\hat{\boldsymbol{X}},\hat{\boldsymbol{X}})+\hat{\boldsymbol{\Sigma}}_{\boldsymbol{y}}$ with $\hat{\boldsymbol{C}}_{\boldsymbol{y}\oplus\boldsymbol{y}'}$, and the test--training cross-covariance $\hat{\boldsymbol{K}}(\hat{\boldsymbol{X}}_*,\hat{\boldsymbol{X}})$ with $\hat{\boldsymbol{K}}_{*,\,\boldsymbol{y}\oplus\boldsymbol{y}'}$.

\subsection{Specifying the Gaussian Process Prior}

At this stage, the posterior GP is fully specified up to the choice of covariance kernel. 
The kernel function encodes the prior assumptions that govern the smoothness, correlation structure, and uncertainty of the free energy profile, and are selected based on known physical behaviors of the underlying latent function.
free energy profiles are generally assumed to be continuous and at least twice differentiable ($C^2$) in physically meaningful collective variables (such as distances, angles, coordination numbers, etc), which is sufficient for defining mean forces (first derivatives) and local curvature (second derivatives).

Here we model the free energy as a smooth, or $C^\infty$, function that is continuously differentiable up to any order derivative with the squared-exponential (or radial basis function) kernel,
\begin{equation}
k_{\text{SE}}(\boldsymbol{\xi},\boldsymbol{\xi}') = w^2 \exp\!\left( -\frac{\left| \boldsymbol{\xi} - \boldsymbol{\xi}' \right|^2}{2 \ell^2} \right).
\end{equation}
\noindent where $\ell$ is a length scale and $w$ is the kernel amplitude.
This enforces a stronger regularity than is strictly required for defining mean forces or local curvature ($C^2$). 
Other kernels for GP regression tasks are well described in Rasmussen and Williams \cite{rasmussen_gaussian_2006} and were not investigated in this work.

\paragraph{A Note on Noise}\label{sec:noise}

Noise is typically estimated directly from molecular simulation statistics, for example using histogram-based variance estimators. 
Implicit in this approach is the assumption that the GP prior is sufficiently expressive to represent the free energy profile, such that any residual discrepancy between the observations and the GP model arises solely from statistical uncertainty in the simulation. 
In practice, however, residuals may also reflect model mismatch arising from limitations of the chosen GP prior. 
Consequently, fixing the noise from simulation statistics alone attributes all model-data discrepancy to sampling uncertainty, potentially leading to pathological hyperparameter inference and misleading uncertainty estimates.

We address this limitation by treating the noise term as an unknown hyperparameter and inferring it directly from the data. 
Under this interpretation, the noise represents an effective measure of data–model mismatch rather than pure measurement error, incorporating simulation uncertainty, temporal and spatial correlations not represented explicitly in the diagonal noise approximation, numerical errors, and fluctuations in the data that the GP regression may not be fully equipped to explain. 
As we will show, this simple reinterpretation prevents the optimizer from converging to non-physical hyperparameter configurations and yields uncertainty estimates that respond appropriately to data acquisition.

\subsection{Generalized Hyperposterior Learning}

Previous work suggested that free energy profile reconstruction is relatively insensitive to hyperparameter selection, leading to the use of fixed hyperparameters chosen \textit{a priori}.
In practice this approach can be well-justified for estimating the profile mean, particularly if one already has strong prior knowledge of the behavior of the underlying function. 
However, selecting hyperparameters based on expert knowledge effectively fixes predictive uncertainty, preventing adaptive response of the UQ to new data.

We therefore relax the assumption that the GP hyperparameters can be specified \textit{a priori}. 
A common alternative is to optimize them from observed data by maximizing the GP marginal likelihood (or, more generally, the posterior over hyperparameters), yielding a \textit{maximum a posteriori} (MAP) estimate. 
Although MAP estimation is computationally efficient compared to sampling the full hyperposterior with Monte Carlo methods, it conditions the predictive distribution on a single set of hyperparameters and therefore neglects their uncertainty. 
This approximation can lead to underdispersed predictive uncertainty, particularly when the available data are limited \cite{rasmussen_gaussian_2006}.

Rather than conditioning the GP on a single optimized set of hyperparameters, we construct and sample the full hyperparameter distribution, yielding a hierarchical Bayesian model. 
The target distribution is defined by a physically motivated hyperprior combined with a leave-one-out (LOO) cross-validation pseudo-likelihood \cite{sundararajan_predictive_2001}, which has previously been shown to improve uncertainty quantification in computational chemistry applications \cite{bartok_improved_2022, shanks_accelerated_2024}. 
In the present work, the LOO objective also produced more informative uncertainty estimates than the standard GP marginal likelihood (Supporting Information Section~\ref{si:mlvsloo}).

Because the LOO objective is not a proper likelihood, the resulting distributions are formally pseudo-posteriors (or generalized posteriors) rather than exact Bayesian hyperposteriors. 
For simplicity, we retain the term "hyperposterior" throughout the remainder of the manuscript. 
Complete specifications of the GP mean functions, covariance kernels, and hyperpriors used in each analysis are provided in Supporting Information Section~\ref{si:hypers}.

\paragraph{Hamiltonian Monte Carlo Implementation}

Hamiltonian Monte Carlo with the No-U-Turn Sampler (HMC-NUTS), as implemented in \texttt{Pyro}~\cite{bingham_pyro_2019}, was used for hyperposterior sampling. 
The target density was the sum of the log hyperprior and the LOO pseudo-likelihood of the joint histogram-derivative GP model. 
Positive kernel and noise hyperparameters were sampled in unconstrained log space and transformed back by exponentiation within the probabilistic model, the so-called "log trick". 
For each proposed hyperparameter state, the full joint covariance matrix of the function and derivative observations was assembled in \texttt{PyTorch}, regularized with a small diagonal "jitter" term for numerical stability ($10^{-8}$), and factorized via Cholesky decomposition to evaluate the target log density. 
Gradients required for leapfrog integration were obtained by automatic differentiation through the GP matrix operations.

An initial burn-in phase of 500 steps was used to adapt the HMC step size and mass matrix, followed by a production phase of 1000 retained samples, which was sufficient for stable estimation of free energy observables with four independent NUTS chains (see Supporting Information Section~\ref{si:hmcdiagnostics}). 
Chains were initialized from dispersed random starting points and run in parallel. 
Convergence was assessed using standard multi-chain Markov chain Monte Carlo diagnostics extracted from the \texttt{Pyro} sampler output, including the split potential scale reduction factor ($\hat{R}$), effective sample size (ESS), divergence counts, and chain-resolved trace plots and marginal posterior histograms for each sampled hyperparameter. 
We defined convergence when $\hat{R} < 1.01$ and bulk and tail ESS exceeded 200 for all parameters (with no persistent divergences). 
We note that standard HMC convergence diagnostics do not necessarily ensure complete exploration of the hyperposterior and may fail to detect unvisited modes in multimodal parameter spaces, but this effect is mitigated by initiating chains in different regions of the hyperparameter space.

\paragraph{Hyperposterior Uncertainty Propagation}

The propagation of hyperparameter uncertainty can be understood through the GP predictive distribution at test inputs $\hat{\boldsymbol{X}}_*$ conditioned on the hyperparameters,
\begin{equation}
    p(\boldsymbol{f}_* \mid \boldsymbol{y}, \boldsymbol{\theta})
    = \mathcal{N}\!\big( \boldsymbol{\mu}_{\boldsymbol{\theta}},\, \hat{\boldsymbol{\Sigma}}_{\boldsymbol{\theta}} \big),
\end{equation}
where $\boldsymbol{\mu}_{\boldsymbol{\theta}} \in \mathbb{R}^{N_*}$ and $\hat{\boldsymbol{\Sigma}}_{\boldsymbol{\theta}} \in \mathbb{R}^{N_*\times N_*}$ are the GP posterior mean and covariance. 
To obtain the hyperparameter-marginalized predictive distribution, we marginalize the conditional Gaussian process predictive distribution over the generalized hyperposterior, $p(\boldsymbol{\theta} \mid \boldsymbol{y})$,
\begin{equation}
    p(\boldsymbol{f}_* \mid \boldsymbol{y})
    = \int p(\boldsymbol{f}_* \mid \boldsymbol{y}, \boldsymbol{\theta})\, p(\boldsymbol{\theta} \mid \boldsymbol{y})\, d\boldsymbol{\theta},
\end{equation}
which yields a mixture of Gaussian distributions.
Although this hierarchical predictive distribution is not Gaussian in general, its first and second moments are given in closed form by the laws of total expectation and total variance \cite{lalchand_approximate_2020},
\begin{equation}
    \bar{\boldsymbol{\mu}}
    = \mathbb{E}[\,\boldsymbol{f}_* \mid \boldsymbol{y}\,]
    = \mathbb{E}_{\boldsymbol{\theta} \mid \boldsymbol{y}}\!\left[\, \boldsymbol{\mu}_{\boldsymbol{\theta}} \right],
    \label{eq:fb-mean}
\end{equation}
\begin{equation}
\begin{aligned}
    \bar{\boldsymbol{\Sigma}}
    &= \mathrm{Cov}(\boldsymbol{f}_* \mid \boldsymbol{y}) \\
    &= \mathbb{E}_{\boldsymbol{\theta} \mid \boldsymbol{y}}\!\big[ \, \hat{\boldsymbol{\Sigma}}_{\boldsymbol{\theta}} \big]
     \;+\;
       \mathrm{Cov}_{\boldsymbol{\theta} \mid \boldsymbol{y}}\!\left(\boldsymbol{\mu}_{\boldsymbol{\theta}}\right),
\end{aligned}
\label{eq:fb-cov}
\end{equation}
or equivalently,
\begin{equation}
\begin{aligned}
    \bar{\boldsymbol{\Sigma}}
    &= \mathbb{E}_{\boldsymbol{\theta} \mid \boldsymbol{y}}\!\big[ \, \hat{\boldsymbol{\Sigma}}_{\boldsymbol{\theta}} \big]
     + \mathbb{E}_{\boldsymbol{\theta} \mid \boldsymbol{y}}\!\left[\, \boldsymbol{\mu}_{\boldsymbol{\theta}}\boldsymbol{\mu}_{\boldsymbol{\theta}}^{\!\top} \right]
     - \bar{\boldsymbol{\mu}}\,\bar{\boldsymbol{\mu}}^{\!\top}.
\end{aligned}
\label{eq:fb-cov-expanded}
\end{equation}

Given $M$ samples $\{\boldsymbol{\theta}^{(m)}\}_{m=1}^M$ from $p(\boldsymbol{\theta} \mid \boldsymbol{y})$, we approximate Eqs.~\eqref{eq:fb-mean}–\eqref{eq:fb-cov-expanded} via,
\begin{equation}
    \bar{\boldsymbol{\mu}}
    \;\approx\;
    \frac{1}{M}\sum_{m=1}^M \boldsymbol{\mu}_{\boldsymbol{\theta}^{(m)}},
\end{equation}
\begin{equation}
\begin{aligned}
    \bar{\boldsymbol{\Sigma}}
    \;\approx\;&\;
    \frac{1}{M}\sum_{m=1}^M \hat{\boldsymbol{\Sigma}}_{\boldsymbol{\theta}^{(m)}} +
    \frac{1}{M}\sum_{m=1}^M 
        \big( \boldsymbol{\mu}_{\boldsymbol{\theta}^{(m)}} - \bar{\boldsymbol{\mu}} \big)
        \big( \boldsymbol{\mu}_{\boldsymbol{\theta}^{(m)}} - \bar{\boldsymbol{\mu}} \big)^{\!\top}.
\end{aligned}
\label{eq:fb-cov-mc}
\end{equation}

The first term represents the average posterior variance conditioned on $\boldsymbol{\theta}$, while the second term accounts for the uncertainty in the GP hyperparameters.
This hyperposterior uncertainty propagation step therefore propagates uncertainty arising from both the GP prediction and the inferred hyperparameters into the reconstructed free energy profile.

Note, however, that the interpretation of this uncertainty is subtle because only free energy differences are physically observable, leaving the free energy profile defined only up to an arbitrary additive constant. 
The resulting predictive covariance therefore reflects uncertainty only in the relative free energy profile, independent of the arbitrary choice of energy zero. 
In practical terms, this uncertainty characterizes the physically meaningful features of the landscape, including the relative depths of metastable states and the free energy barriers separating them.

\subsection{Evaluation of Uncertainty under Data Ablation}

To assess how uncertainty changes as a function of the amount of available data, we performed free energy UQ while systematically knocking-out umbrella windows and trajectory time. 
As a test system, we studied umbrella sampling data of a nonaarginine peptide (R9) interacting with a lipid bilayer. 
The CV was defined as the distance between the centers of mass of the membrane and peptide normal to the membrane surface.
The dataset consists of $25$ umbrella windows, each constrained by a harmonic potential with $\sim 500$ ns long trajectories.
Simulation details are described elsewhere \cite{baxova_direct_2026}.

The four methods presented are: (a) umbrella integration (UI) with block averaging (methodology in Supporting Information Section~\ref{si:uiblock}), (b) a fixed hyperparameter GP regression model with simulation derived noise variances, (c) our hierarchical GP model with optimized hyperparameters (MAP estimation), and (d) the same hierarchical GP but with hyperposterior uncertainty propagation.

After removing equilibration timesteps, the trajectory length was cut from the beginning up to a selected percentage of the converged simulation run, from $4$ to $100$ \% ($T = [4, 6.3, 10, 16, 25, 40, 60, 80, 90, 100])$. 
The number of windows were varied from $3$ to the maximum available of $25$ ($W = [3,4,6,8,10,13,16,19,22,25]$). 
A $T \times W$ grid was constructed for a total of 100 UQ calculations. 

Five independent calculations were performed for each grid evaluation. 
One used an evenly spaced subset of umbrella windows, while the remaining four were randomly selected subsets of the available 25 windows. 
RMSE with respect to a full dataset WHAM reference and predictive standard deviation were calculated for each reconstruction and averaged over these five computations.
Averaging over window selection was introduced to mitigate bias associated with the window selection process, which in general may be different depending on the Bayesian optimization (or equivalent) decision-theoretic scheme employed.
Heatmaps of the average RMSE and standard deviations were constructed to visualize the relationship between error and predictive uncertainty.

\begin{figure*}[h]
    \centering
    \includegraphics[width=\textwidth]{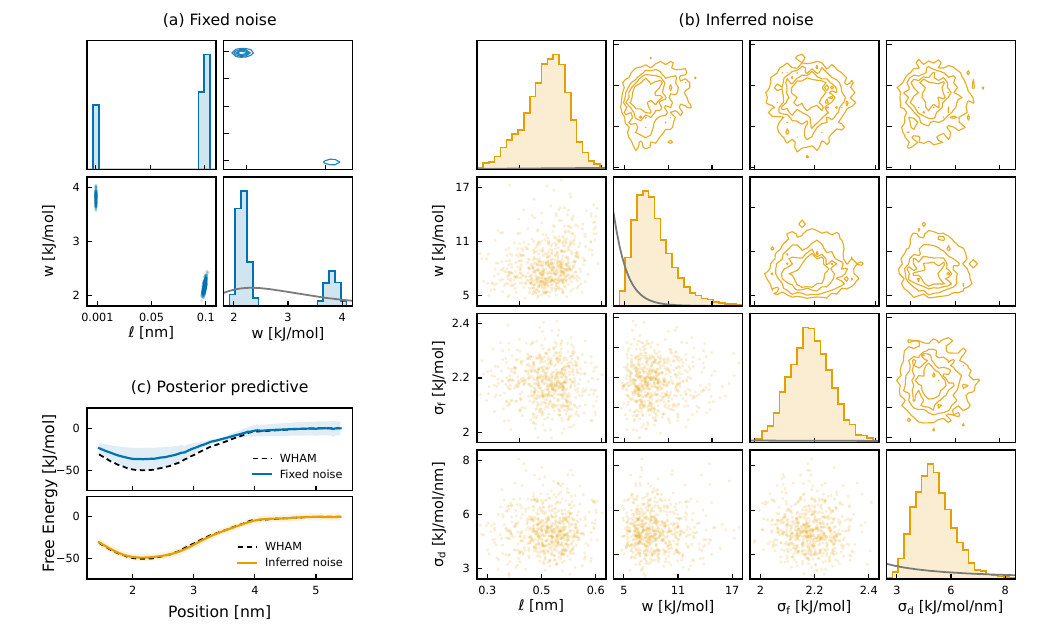}
    \caption{Comparison between fixed and inferred noise hyperparameter models. (a) Corner plot of fixed noise hyperposterior with marginal hyperpriors (grey line). (b) Corner plot of inferred noise hyperposterior. Raw HMC-NUTS samples below diagonal and sample contour map above diagonal. (c) Posterior predictive surfaces comparing the fixed and inferred noise models with shaded region given by $\pm \sigma$.}
    \label{fig:noise_inference}
\end{figure*}

\section{Results}

\subsection{Noise as an Uncertain Hyperparameter} 

GP-based free energy methods typically estimate histogram and derivative noise variances from simulation statistics (Supporting Information Eqs.~\eqref{si:eq_histnoise} and \eqref{si:eq_dernoise}) and condition the GP posterior on these fixed values.
However, this implicitly assumes that the prior is sufficiently expressive to fit the training data such that any remaining discrepancy between the observations and GP can be attributed to statistical uncertainty in MD sampling alone.
In general, this assumption need not hold, since the posterior free energy profile is coupled to both the MD data and the GP prior.
Treating the noise variance as known therefore forces the GP to explain any residual data-model mismatch through its remaining hyperparameters, collapsing the hyperposterior into unlikely regions of the hyperparameter space.

The consequences of this approximation are shown in generalized hyperposterior MAP summarized in Table~\ref{tab:noise_hyperparameter_map} and posterior distributions visualized in Figure~\ref{fig:noise_inference}. 
For a fixed noise model, the hyperposterior over the $\ell$ and $w$ hyperparameters is multimodal, with a MAP estimate of $\ell = 0.10$ nm and $w = 1.88$ kJ/mol (Figure~\ref{fig:noise_inference}a). 
The second mode collapses to an unrealistically short $\ell=$0.01 nm accompanied by an inflated kernel amplitude $w=$3.95 kJ/mol.
The consequences of this bimodality are clear in the sampled GP posterior, with an RMSE of 9.20 kJ/mol and predictive standard deviation of 7.97 kJ/mol.
Small discrepancies between the plotted posterior histograms and the reported MAP estimates arise because the MAP values were obtained by direct optimization of the joint objective, rather than by estimating marginal modes from binned histograms.
The histogram shapes therefore provide a visual summary of the sampled marginal distributions, but their bin-dependent maxima should not be interpreted as the joint MAP.

\begin{table}[h]
\centering
\caption{LOO MAP hyperparameter estimates for fixed- and inferred-noise GP models.}
\vspace{4pt}
\label{tab:noise_hyperparameter_map}
\begin{tabular}{lrrrr}
\hline
Noise model & $\ell$ & $w$ & $\sigma_\mathrm{f}$ & $\sigma_\mathrm{d}$ \\
\hline
Fixed    & 0.10 & 1.88 & --   & --   \\
Inferred & 0.48 & 7.51 & 2.18 & 4.94 \\
\hline
\end{tabular}
\end{table}

Relaxing the delta-function priors on the nuisance parameters $\sigma_\mathrm{f}$ and $\sigma_\mathrm{d}$ acknowledges that statistical uncertainty from the enhanced sampling MD is not the only source of discrepancy between the observations and the GP model. 
As shown in Figure~\ref{fig:noise_inference}b, jointly inferring these quantities removes the pathological short-length-scale mode and yields a well-behaved hyperposterior centered around $\ell = 0.48$ nm and $w = 7.2$ kJ/mol. 
The inferred values of $\sigma_\mathrm{f}$ and $\sigma_\mathrm{d}$ are larger than those predicted from simulation statistics alone, indicating that the model attributes part of the residual discrepancy to limitations of the GP representation rather than forcing it into the kernel hyperparameters. 
This change also resolves the deterioration in predictive RMSE and average standard deviation, with values of 1.59 and 4.12 kJ/mol, respectively.
Modeling the histogram and derivative noise variances as inferred hyperparameters therefore appears to correct hyperparameter identifiability and uncertainty propagation pathologies from the generalized hyperposterior. 
This idea represents the principal conceptual departure of our framework from prior work. 

\subsection{Uncertainty Evolution with Data Acquisition}

\begin{figure*}[t!]
    \centering
    \includegraphics{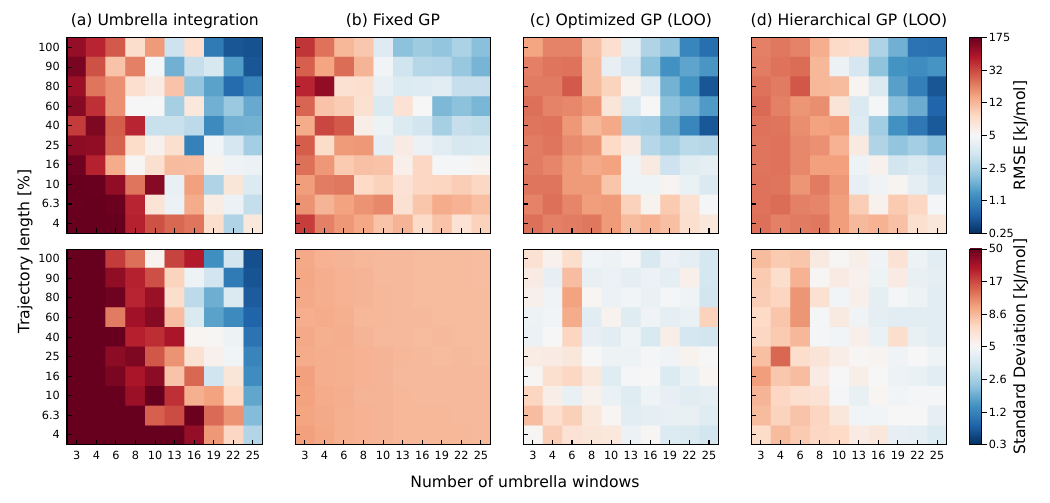}
    \caption{Data ablation heatmap grids. Columns represent different free energy profile reconstruction and error estimation methods: (a) Umbrella Integration (UI) with the block averaging, (b) GP with fixed hyperparameters, (c) optimized hierarchical GP model with LOO pseudo-likelihood and (d) sampled hierarchical GP model with LOO pseudo-likelihood. The first row shows the mean RMSE with respect to the reference WHAM prediction, while the second row shows the average predictive standard deviation in the given model. Color bars are shared within each row and shown on a logarithmic scale. White denotes a chemical accuracy threshold (RMSE = 5 kJ/mol, top row; average SD = 5 kJ/mol, bottom row).}
    \label{fig:ablation_grid}
\end{figure*}

An important property of reliable UQ methods is that the estimated uncertainty should be large when the prediction is poor, and only become small when the prediction is accurate.
This behavior ensures that we are close to an optimal min-max problem, where we minimize computational time while still obtaining accurate and trustworthy predictions.
We evaluated four UQ methods on our R9 umbrella sampling dataset to assess their ability to reproduce this desirable behavior by knocking out umbrella windows and trajectory time.
Heatmaps of average RMSE and standard deviation for these data ablation grids are shown in Figure~\ref{fig:ablation_grid}.

All methods considered show the expected trend that increasing the amount of simulation data improves reconstruction accuracy, resulting in lower average RMSE values (top row of Figure~\ref{fig:ablation_grid}). 
Furthermore, the RMSE of the free energy prediction was excellent in the high-data limit for all tested methods.
For this particular system, increasing the number of umbrella windows provided a larger improvement than extending the trajectory length within individual windows. 
However, this observation is system dependent and should not be interpreted as a general rule.

The bottom row of Figure~\ref{fig:ablation_grid} shows heatmaps of the average predictive standard deviation for each method, where red indicates values greater than 5 kJ/mol and blue values below this threshold. 
For short trajectories and low window counts, UI with block averaging produces large standard deviations, reaching values up to 265 kJ/mol. 
Large uncertainty estimates persist even in regions of the grid corresponding to 8-22 umbrella windows, where the free energy profile mean is already within chemical accuracy. 
As additional data are introduced, the standard deviation estimates collapse abruptly toward near-zero values. 
This behavior suggests that the uncertainty estimates are not well-correlated with the reconstruction error and therefore provide a misleading picture of convergence.
Specifically, this method appears underconfident in that it will begin to achieve high accuracy (low RMSE) before the predictive uncertainties converge.

The fixed-hyperparameter GP (fixed GP) model produces a different pathology. 
The uncertainty remains nearly unchanged regardless of the amount of available data (8.3--9.2 kJ/mol), as evidenced by the essentially uniform standard deviation heatmap. 
Consequently, the predictive uncertainty becomes largely decoupled from the information content of the simulations and cannot reliably indicate whether additional sampling has improved the reconstruction.

In contrast, the hierarchical GP models provide uncertainty estimates that evolve consistently with the amount of available data. 
The optimized GP (Figure~\ref{fig:ablation_grid}c) is overconfident in the low-data setting, with a predictive standard deviation approximately 18\% lower, on average, than the hyperposterior-propagated model (Supporting Information Section~\ref{si:hyperopt}). 
Propagating hyperparameter uncertainty with the hierarchical GP corrects this overconfidence (Figure~\ref{fig:ablation_grid}d). 
This comparison suggests that uncertainty associated with the GP hyperparameters contributes meaningfully to the overall predictive uncertainty.

\subsection{Propagating Free Energy Uncertainty}

Access to a probability distribution enables principled uncertainty propagation to downstream quantities-of-interest (QoIs). 
Consequently, uncertainties in the simulation data, model assumptions, and GP hyperparameters can be propagated directly to these observables, yielding predictive distributions rather than point estimates. 

\begin{figure}
    \centering
    \includegraphics[width=1.0\linewidth]{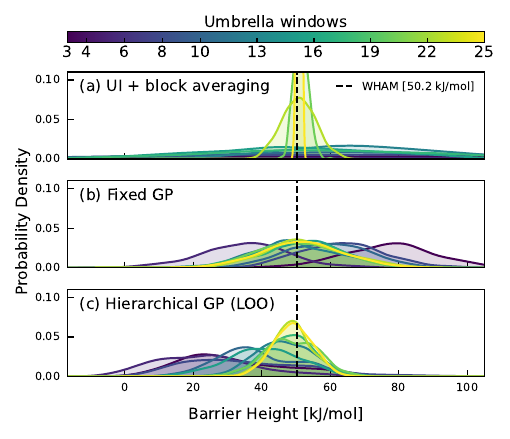}
    \caption{Barrier-height distributions for three PMF estimation methods as a function of sampling window count (100\% trajectory length). Distributions correspond to the free energy barrier height, ($\Delta A^\ddagger = \max(f)-\min(f)$), estimated from 500 samples at each window count and colored from 3 (purple) to 25 (yellow) windows. The vertical dashed line indicates the WHAM reference barrier height.}
    \label{fig:QoI}
\end{figure}

Figure \ref{fig:QoI} shows uncertainty propagation to the free energy barrier height of the R9-membrane interaction as a function of the number of umbrella windows. 
The barrier-height distributions reveal the same qualitative behavior observed for the free energy profiles. 
UI with block averaging, in line with the previous heatmap, produces broad barrier-height distributions for low window counts, reflecting substantial uncertainty in the reconstructed surface. 
The distribution remains broad until a nearly complete dataset is available, at which point it abruptly collapses to an almost delta-function-like distribution centered on a single barrier value.

For the fixed GP, the width of the barrier-height distribution remains nearly unchanged as additional data are acquired. 
Although the mean barrier estimate gradually approaches the reference value, the associated uncertainty does not evolve with the amount of information provided by the simulations. 
Consequently, the uncertainty estimate offers little indication of whether the calculation has converged.

In contrast, the hierarchical GP exhibits a smooth and physically sensible convergence behavior. 
The barrier-height distribution systematically narrows around the reference value while remaining appropriately broad in undersampled regimes. 
This indicates that the model is able to adapt its uncertainty estimates to the amount of available information, yielding both accurate barrier predictions and reasonable uncertainty estimates throughout the convergence process.

\subsection{Extended Lagrangian Metadynamics Benchmark}

Although the preceding results focused on an umbrella sampling example, the hierarchical GP formulation is applicable to data obtained from any enhanced sampling method. 
To illustrate this, we applied our framework to extended Lagrangian metadynamics simulations of phenol interacting with a lipid bilayer. 
As in the previous case, the CV is defined as the distance between the centers of mass of the membrane and the phenol molecule normal to the membrane surface.
The primary distinction from umbrella sampling is that histogram and derivative observations are not naturally partitioned into independent umbrella windows, but must instead be constructed by discretizing the metadynamics trajectory. 
This additional preprocessing step introduces another layer of modeling assumptions, making metadynamics a useful test of the generality of the framework. 
Details of the molecular simulations and discretization procedure are provided in Supporting Information Section~\ref{si:metad}.

\begin{figure}
    \centering
    \includegraphics{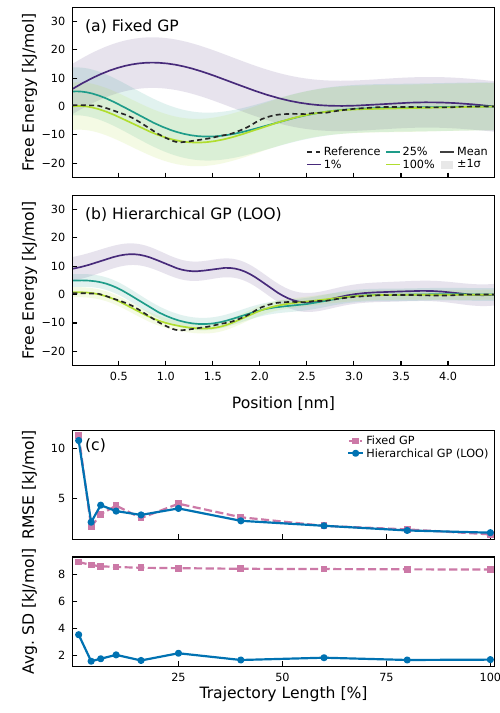}
    \caption{Metadynamics GP-based free energy reconstruction as a function of trajectory length. Fixed GP and hierarchical GP posterior predictive means are shown relative to the WHAM reference for selected trajectory fractions; shaded regions denote \(\pm \sigma\). Bottom panels show the RMSE and average predictive standard deviation as a function of retained trajectory length for the fixed GP and hierarchical GP models.}
    \label{fig:metad}
\end{figure}

In analogy to the umbrella-sampling analysis, we examined how uncertainty changes with the retained length of the metadynamics trajectory.
The results for the fixed and hierarchical GP are shown in Figure~\ref{fig:metad}.
The fixed GP behaved similarly to the umbrella sampling case, showing improving RMSE with additional trajectory data, decreasing from 11.30 to 1.41 kJ/mol, but nearly constant average predictive standard deviation, varying only from 8.89 to 8.33 kJ/mol. 
Thus, the fixed GP model did not adapt its uncertainty scale appreciably as additional metadynamics data were included.

In contrast, the hierarchical GP produced uncertainty estimates that changed with the amount of available data and remained broadly correlated with reconstruction accuracy.
For the hierarchical GP, the average predictive standard deviation decreased rapidly from 3.56 kJ/mol at 1\% of the trajectory to approximately 1.6--2.2 kJ/mol over the remaining trajectory lengths. 
Over the same range, the RMSE decreased from 10.77 kJ/mol at 1\% trajectory length to 1.58 kJ/mol for the full trajectory. 
The average predictive standard deviation was correlated with the predictive mean error, with a Pearson correlation coefficient of \(\rho=0.95\).

These results suggest that, despite the additional modeling assumptions required to discretize metadynamics trajectories, the hierarchical GP framework can still provide reliable UQ beyond umbrella sampling. 
The fixed GP remains a useful baseline, but its uncertainty evolution is substantially less responsive to the amount of metadynamics data.

\section{Discussion}

The results from the previous sections reinforce a simple but important lesson: uncertainty quantification for free energy profiles is inseparable from the assumptions of the underlying statistical model. 
Consequently, strong assumptions about GP hyperparameters inevitably propagate into the predictive uncertainties.
In this study, we showed that the treatment of noise, the choice of likelihood (or pseudo-likelihood) objective, and uncertainty in the GP hyperparameters can alter predictive uncertainty estimation for free energy reconstruction, even in simple one-dimensional problems. 
Our results suggest that reliable free energy profiles, accompanied by meaningful uncertainty estimates, can often be reconstructed from substantially less data than is required by conventional free energy reconstruction methods. 
This may be particularly valuable for \textit{ab initio} molecular dynamics (AIMD), where the computational expense of generating sufficiently converged free energy landscapes remains a major practical limitation.

In the context of automated free energy workflows, omitting sources of uncertainty becomes particularly problematic because the simulation data are sparse, noisy, and highly dependent on sampling protocol. 
We showed that generalized hierarchical Bayesian inference can mitigate this problem by propagating uncertainty in the GP hyperparameters directly into the predictive distribution. 
As a result, the model is able to adapt to the available information and provides a more reliable measure of convergence.
This behavior was demonstrated for umbrella sampling and extended Lagrangian metadynamics examples, but in principle is applicable to any free energy method, including the accelerated weighted histogram~\cite{lindahl_accelerated_2014}, alchemical decoupling~\cite{york_modern_2023}, and multistate Bennet acceptance ratio \cite{ding_2024} methods.

The hierarchical GP framework has its limitations, perhaps the most significant being computational cost. 
Unlike conventional GP-based free energy reconstruction, which requires only a single evaluation of the predictive posterior at fixed hyperparameters, this approach requires repeated assembly and factorization of a joint covariance matrix for each HMC sample of the hyperposterior. 
For the one-dimensional problems considered here, this additional cost is negligible compared to that of generating well-sampled MD trajectories, but could become a limiting factor for high-dimensional surfaces.
Furthermore, if one needs to increase the flexibility of the GP prior (which typically introduces additional hyperparameters), the computational burden associated with Bayesian inference will also increase. 
Developing scalable inference strategies for richer GP models therefore remains an important area for future work.

Beyond computational considerations, the reliability of any Bayesian inference procedure is ultimately limited by the assumptions encoded in the prior model. 
In the present work, we considered only stationary covariance kernels, which assume that correlations depend solely on distance in CV space and not on location. 
While this assumption is common, there is no physical reason why a free energy profile should exhibit uniform correlation structure throughout an entire CV domain. 
Regions near metastable basins, transition states, and steep free energy barriers may possess fundamentally different characteristic length scales and smoothness properties. 
A natural extension of the present framework would therefore be the use of non-stationary kernels, either in parametric form \cite{sullivan_physics-informed_2025} or fully nonparametric form \cite{heinonen_non-stationary_2016}. 
Such models would permit spatially varying correlation structures and potentially provide more realistic uncertainty estimates, albeit at the cost of a substantially more complex hyperposterior inference problem.

At the same time, a GP-based representation of free energy profiles creates opportunities for integration with a broader ecosystem of Bayesian methodologies in computational chemistry. 
GPs have already been successfully applied to scattering analysis \cite{sullivan_physics-informed_2025}, force field optimization \cite{shanks_accelerated_2024, shanks_bayesian_2024, fan_charge_2025, kostal_bayesian_2026}, and machine-learning interatomic potentials \cite{bartok_gaussian_2010, shanks_transferable_2022, shanks_experimental_2025}. 
Representing free energy profiles within the same probabilistic framework raises the possibility of combining information from multiple experimental and computational sources into unified Bayesian models. 
Such approaches could enable consistency checks between scattering data, electronic structure calculations, force field models, and free energy calculations, while also providing a mechanism for propagating uncertainty between different levels of theory.

\section{Conclusions}

We evaluated several strategies for quantification of uncertainties for free energy calculations and found that generalized hierarchical GP inference was the only approach whose uncertainty estimates consistently adapted to the amount and quality of the available simulation data. 
In contrast, existing methods exhibited systematic failure modes by conditioning on fixed modeling assumptions, including fixed hyperparameters and noise models that attribute all uncertainty to statistical sampling. 
These results highlight the importance of propagating uncertainty throughout the statistical model rather than conditioning on point estimates or fixed probabilistic assumptions. 
Although important challenges remain—including computational scaling, higher-dimensional collective variable spaces, and more expressive non-stationary covariance models—we conclude that hierarchical GP inference provides a promising foundation for next-generation automated free energy calculations in which uncertainty quantification is treated as a primary objective rather than an afterthought.

\section{Author Information}

\textbf{CRediT Author Contributions}

M. Skorna:
Conceptualization (supporting), Formal analysis (equal), Investigation (equal), Methodology (equal), Software (equal), Validation (equal), Visualization (equal), Writing – original draft (equal)

A. Gottfried:
Conceptualization (supporting), Formal analysis (supporting), Investigation (supporting), Methodology (supporting), Writing – original draft (supporting)

Z. Janackova:
Conceptualization (supporting), Data curation (lead), Investigation (supporting), Methodology (supporting), Visualization (supporting), Writing – original draft (supporting), Writing – review \& editing (supporting)

K. Baxova:
Data curation (supporting), Methodology (supporting)

P. Jungwirth:
Funding acquisition (lead), Project administration (lead), Resources (lead), Supervision (equal), Writing – review \& editing (supporting)

B. L. Shanks:
Conceptualization (lead), Formal analysis (equal), Investigation (equal), Methodology (equal), Project administration (supporting), Software (equal), Supervision (equal), Validation (equal), Visualization (equal), Writing – original draft (equal), Writing – review \& editing (lead) 

\textbf{Notes}

The authors declare no competing financial interest.

\section{Data Availability}

The code and datasets used in this work are provided on GitHub at \url{https://github.com/brennonshanks/freeGP} and Zenodo \url{https://zenodo.org/uploads/21358802}, respectively.
A tutorial folder in the GitHub repo allows users to try the code on an artificial system and the repo README provides guidance on adapting umbrella sampling and metadynamics data into a suitable format to work with our code base.
Additional information is available from the corresponding authors upon reasonable request.

\appendix
\section{Umbrella Integration with Block Averaging}\label{si:uiblock}

After equilibration removal and trajectory truncation, each umbrella trajectory was divided into 10 contiguous blocks. 
For each umbrella window \(i\) and block \(b\), a Gaussian approximation to the biased coordinate distribution was used to estimate the block-specific mean-force contribution, \(A'_{i,b}(x)\).
The window-specific mean-force estimate was then obtained by averaging over the block estimates,
\begin{equation}
\overline{A'_i}(x)=\frac{1}{B}\sum_{b=1}^{B}A'_{i,b}(x),
\end{equation}
where \(B=10\) is the number of blocks. 
The uncertainty in the window mean force was estimated from the squared standard error (SE),
\begin{equation}
\mathrm{SE}^2\!\left[A'_i(x)\right]=\frac{\mathrm{Var}_b\!\left[A'_{i,b}(x)\right]}{B}.
\end{equation}

The window-specific mean forces were combined using the standard UI weights,
\begin{equation}
\overline{A'}(x)=\sum_i\tilde{w}_i(x)\, \overline{A'_i}(x),
\end{equation}
where \(\tilde{w}_i(x)\) are the normalized UI weights. 
Assuming independent uncertainties between umbrella windows, the variance of the combined mean force was approximated as
\begin{equation}
\mathrm{Var}\!\left[\overline{A'}(x)\right]\approx\sum_i\tilde{w}_i(x)^2\,\mathrm{SE}^2\!\left[A'_i(x)\right].
\end{equation}

The PMF was obtained by trapezoidal quadrature of the mean-force profile,
\begin{equation}
A(x_j)=A(x_0)+\sum_{k=1}^{j}\frac{x_k-x_{k-1}}{2}\left[\overline{A'}(x_{k-1})+\overline{A'}(x_k)\right].
\end{equation}

The pointwise mean-force variance was propagated through the same trapezoidal integration procedure using a running sum to estimate the within-replicate free energy variance. 
This propagation assumes that mean-force uncertainties at different grid points are independent and therefore neglects covariance between neighboring mean-force estimates. 
For each data condition, multiple independent window selections were analyzed. 
The total UI uncertainty was computed as the sum of the average within-replicate variance and the variance across replicate window selections, consistent with the law of total variance.

\section{Hyperposterior Specification}\label{si:hypers}

The GP was assigned a squared-exponential covariance function prior,
\begin{equation}
k_{\text{SE}}(\boldsymbol{\xi}, \boldsymbol{\xi}') = w^2 \exp\!\left( -\frac{\left| \boldsymbol{\xi} - \boldsymbol{\xi}' \right|^2}{2 \ell^2} \right).
\end{equation}
where $\ell$ is the correlation length and $w$ is the kernel amplitude. 
All positive hyperparameters were sampled in natural-log space. 
Expressing distances in nm and free energies in kJ/mol, the default independent hyperpriors were
\[
\begin{aligned}
\log \ell &\sim \mathcal{N}(\log 4,\,1^2),\\
\log w &\sim \mathcal{N}(1,\,0.5^2),\\
\log \sigma_\mathrm{f} &\sim \mathcal{N}(0.5,\,2^2),\\
\log \sigma_\mathrm{d} &\sim \mathcal{N}(0.5,\,2^2).
\end{aligned}
\]
Equivalently, $\ell$, $w$, $\sigma_\mathrm{f}$, and $\sigma_\mathrm{d}$ follow log-normal distributions with medians of 4 nm, 2.72 kJ/mol, 1.65 kJ/mol, and 1.65 kJ/mol/nm, respectively. 
The length-scale prior was intentionally chosen to be weakly informative, allowing the posterior to learn substantially shorter or longer correlation lengths when supported by the data. 
The parameters $\sigma_\mathrm{f}$ and $\sigma_\mathrm{d}$ represent nuisance noise scales whose variances enter the diagonal of the function and derivative observation covariance matrices. 
Their priors were intentionally broad to permit uncertainty arising from both sampling variability and model mismatch. 

Unless otherwise stated, these hyperpriors were combined with the LOO pseudo-likelihood to construct the generalized hyperposterior.

\subsection{Details of the Fixed Hyperparameter GP}

The fixed-hyperparameter reference used the same stationary covariance with $\ell=\pi/2$ nm and $w=4.184\sqrt{10}$ kJ/mol (the heuristic recommendation is $\sqrt{10}$ kcal/mol). 
Although these kernel parameters were fixed, the observation covariance was estimated from each umbrella trajectory. 
After equilibration removal, the CV time series $\mathbf{x}$ in each window was approximated as an autoregressive process of order one. 
This provides an inexpensive approximation to the integrated autocorrelation time. 
The lag-one coefficient was estimated using a weakly regularized
ratio of the lag-one sample autocovariance to the sample variance,
\[
\hat{\phi}=\tanh\left[\frac{\sum_t (x_t-\bar{x})(x_{t+1}-\bar{x})}
{\lambda+\sum_t (x_t-\bar{x})^2}\right],
\]
with $\lambda=0.01$ in the squared coordinate units of the input trajectory.
The regularization shrinks poorly determined estimates toward zero and the $\tanh$ transformation constrains $\hat{\phi}$ to the stationary interval $(-1,1)$. 
The corresponding AR(1) statistical inefficiency and effective
sample size were
\[\tau=\frac{1+\hat{\phi}}{1-\hat{\phi}},\qquad N_{\mathrm{eff}}=\frac{N}{\tau}.
\]
We did not cap $N_{\mathrm{eff}}$ at $N$; therefore, windows with negative lag-one correlation can have $\tau<1$ and $N_{\mathrm{eff}}>N$. 
In practice, this estimator was used only to scale the trajectory-derived observation covariances for the fixed hyperparameter GP and umbrella integration references, and should be regarded as an approximate AR(1)-based correction rather than a full integrated autocorrelation time analysis.

For a histogram bin with probability \(p_i\), uncertainty in the histogram-derived free-energy observation was represented by
\begin{equation}\label{si:eq_histnoise}
    \operatorname{Var}(F_i)=\frac{1}{\beta^2N_{\mathrm{eff}}}\left(\frac{1}{p_i}-1\right).
\end{equation}
The multinomial constraint introduces covariance between bins from the same umbrella window,
\[
\operatorname{Cov}(F_i,F_j)=-\frac{1}{\beta^2N_{\mathrm{eff}}},\qquad i\ne j.
\]
For the derivative observation obtained from an umbrella with force constant $k$ and positional variance $s_x^2$, the estimated variance was
\begin{equation}\label{si:eq_dernoise}
    \operatorname{Var}(\overline{F'})=\frac{k^2s_x^2}{N_{\mathrm{eff}}}.
\end{equation}

\subsection{Hyperprior Sensitivity Analyses}

Bayesian inference is influenced by prior selection, making sensitivity analyses important for assessing robustness. 
Since it is computationally impractical to exhaustively explore all hyperprior choices, we focus on a case-study of the length scale $\ell$.
$\ell$ is a particularly relevant parameter to investigate, because larger length scales regularize non-physical free energy fluctuations by favoring smoother surfaces. 
We therefore examined several length-scale hyperpriors in a data-limited regime (7 windows, 25\% trajectory), where the influence of prior assumptions should be most apparent. 

Four priors were examined: (i) a prior uniform in $\log \ell$, (ii) the default prior $\log \ell \sim \mathcal{N}(\log 4,1^2)$ used throughout this work, and two informative priors centered at $\ell=0.5$ nm with standard deviations of $0.5$ and $0.1$ in log space.
Priors on all remaining hyperparameters were unchanged.

\begin{figure}[ht!]
\centering
\includegraphics{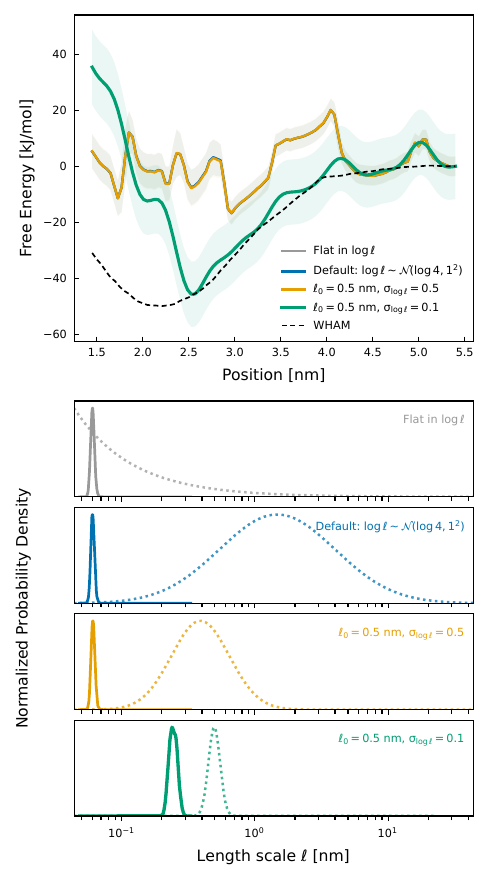}
\caption{
Length-scale hyperprior sensitivity analysis. (a) Posterior free energy profile compared to WHAM on the full umbrella sampling dataset. Shaded regions represent the predictive credible interval of $\pm \sigma$. (b) Normalized $\ell$ priors (dashed) and sampled hyperposterior marginals (solid).}
\label{fig:lengthscale_sensitivity}
\end{figure}

The reconstruction was largely insensitive to the uniform, default, and moderately informative priors (Table~\ref{tab:lengthscale_sensitivity}), suggesting a data-dominated generalized posterior. 
The inferred posterior median length scales differed by less than 1\%, their credible intervals strongly overlapped, and the resulting free energy surfaces were visually indistinguishable. 
Across these three cases, the mean predictive standard deviation ranged from 4.71-4.92 kJ/mol and the RMSE relative to the UI reference ranged from 17.88-17.94 kJ/mol. 

On the other hand, the strongly informative prior centered at 0.5 nm with $\sigma_{\log}=0.1$ shifted the median $\ell$ to 0.244 nm. 
This stronger regularization suppresses short-length-scale structure, but also increased the RMSE relative to UI to 29.98 kJ/mol and increased the mean predictive standard deviation to 11.88 kJ/mol. 
These results suggest that the free energy reconstruction is robust to broad and moderately informative choices of the length-scale hyperprior, but strongly informative priors can significantly impact the reconstruction and thus should be used with care.

\begin{table}[h]
\centering
\caption{
Posterior sensitivity to tested $\ell$ hyperpriors.
}
\label{tab:lengthscale_sensitivity}

\begin{tabular}{lcccc}
\toprule
Prior &
Median $\ell$ (nm) &
RMSE &
SD \\
\midrule

Uniform in $\log \ell$
& 0.0604
& 17.94
& 4.71
\\

$\log \ell \sim \mathcal{N}(\log 4,1^2)$
& 0.0605
& 17.94
& 4.72
\\

$\log \ell \sim \mathcal{N}(\log 0.5,0.5^2)$
& 0.0609
& 17.88
& 4.92
\\

$\log \ell \sim \mathcal{N}(\log 0.5,0.1^2)$
& 0.2443
& 29.98
& 11.88
\\

\bottomrule
\end{tabular}
\end{table}

\section{Hamiltonian Monte Carlo Diagnostics}\label{si:hmcdiagnostics}

To assess convergence of our HMC-NUTS implementation, we systematically varied burn-in period and number of retained posterior samples. 
Four data regimes were considered: an \textit{easy} case consisting of 25 umbrella windows and the full post-equilibration trajectory, a \textit{medium} case with 13 windows and 50\% of each trajectory, a \textit{hard} case with 7 windows and 25\% of each trajectory, and a \textit{super-hard} case with  3 windows and 10\% of each trajectory. 
For each regime, four independent chains were run using 250, 500, or 1000 burn-in steps and 250, 500, or 1000 retained samples per chain, yielding a total of 36 calculations. 
All calculations employed the squared-exponential kernel with inferrable noise hyperparameters, default hyperpriors, and the LOO pseudo-likelihood.

Convergence was assessed using the potential scale reduction factor ($\hat{R}$), effective sample size (ESS), the number of divergent transitions, and visual inspection of chain trace plots and marginal posterior distributions. 
In addition, posterior-derived quantities, including the average predictive standard deviation and free-energy barrier height, were monitored across run lengths to assess sensitivity of physically relevant observables to HMC-NUTS settings.

The chains were well mixed and no divergent transitions were observed. 
For the easy and medium regimes, all calculations yielded excellent convergence diagnostics, with maximum $\hat{R}$ values of 1.0062 and 1.0033, respectively. 
In the hard regime, eight of the nine runs satisfied $\hat{R}<1.01$, with the only exception corresponding to the shortest calculation (250 burn-in steps and 250 retained samples per chain), for which the maximum $\hat{R}$ was 1.0196 and the minimum ESS was 232.

The super-hard regime provided the most stringent test because the limited amount of data weakly constrained the hyperposterior. 
Seven of the nine runs achieved $\hat{R}<1.01$. 
The two failed cases both used only 250 retained samples, yielding maximum $\hat{R}$ values of 1.0107 and 1.0110. 
In contrast, all calculations using 500 or more retained samples satisfied $\hat{R}<1.005$, while all calculations using 1000 retained samples satisfied $\hat{R}<1.002$, irrespective of burn-in length. 
Thus, increasing the number of retained samples per chain is more beneficial than extending the burn-in period beyond 250 steps for weakly identified posteriors.

\begin{figure}[ht!]
    \centering
    \includegraphics{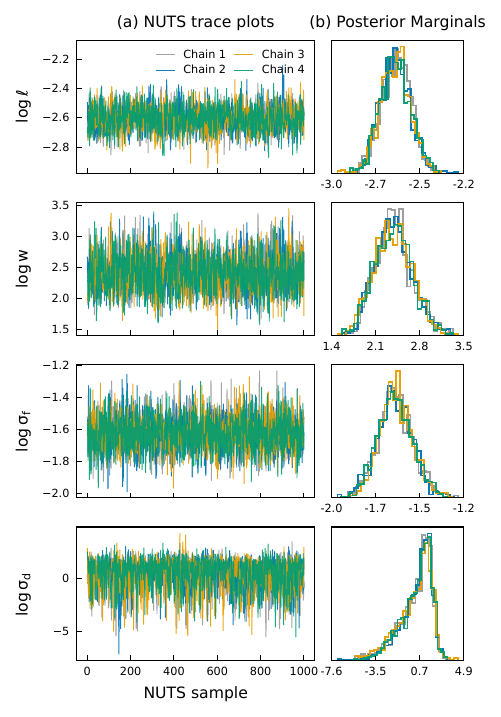}
    \caption{NUTS chain tracking for the super-hard data case (3 windows, 10\% trajectory) exhibits excellent mixing and sampling of the hyperposterior despite limited data.}
    \label{fig:NUTSdiagnostics}
\end{figure}

Production calculations were therefore performed using 500 burn-in steps followed by 1000 retained samples per chain. 
At this setting, all four data regimes exhibited excellent convergence (Table~\ref{tab:nuts_diagnostics}), with a maximum $\hat{R}$ of 1.0024, a minimum ESS of 1479, and no divergent transitions. 
Trace plots showed stationary fluctuations without persistent chain offsets, and marginal posterior distributions exhibited strong overlap across chains (Figure~\ref{fig:NUTSdiagnostics}). 

\begin{table}[h]
\centering
\caption{
NUTS diagnostics and divergences (Divs) using a 500 step burn-in and 1000 retained samples for four chains.
}
\label{tab:nuts_diagnostics}

\begin{tabular}{rrccc}
\toprule
Windows &
Trajectory &
Max $\hat{R}$ &
Min ESS &
Divs \\
\midrule
25 & 1.00 & 1.0008 & 2823 & 0 \\
13 & 0.50 & 1.0012 & 2393 & 0 \\
7  & 0.25 & 1.0024 & 1554 & 0 \\
3  & 0.10 & 1.0018 & 1479 & 0 \\
\bottomrule
\end{tabular}
\end{table}

Posterior-derived observables were similarly insensitive to the precise burn-in and sampling length once standard convergence criteria were satisfied. 
Across all nine burn-in/sample combinations, the coefficient of variation of the mean predictive standard deviation was 1.66\%, 1.31\%, 1.84\%, and 2.93\% for the easy, medium, hard, and super-hard regimes, respectively. 
Corresponding free-energy barrier estimates varied only between 48.57-48.69, 50.40-50.72, 36.84-36.91, and 30.46-30.94 kJ/mol. 
The largest within-regime variation was therefore only 0.49 kJ/mol, observed for the super-hard case.

\section{Sensitivity to Likelihood Objective}\label{si:mlvsloo}

Hyperparameter inference requires defining an objective function for constructing the generalized posterior over the GP hyperparameters. 
The standard Bayesian choice is the marginal likelihood, which evaluates how probable the observed data are under a given set of hyperparameters after integrating over the latent free energy profile. 
However, maximizing the probability of the observed data is not necessarily equivalent to maximizing predictive performance on unseen data. 
Since the primary objective of the present work is reliable uncertainty quantification rather than reconstruction of the mean free energy surface alone, we also considered a leave-one-out (LOO) pseudo-likelihood objective \cite{sundararajan_predictive_2001}. 
Generalized hyperposteriors constructed from both objectives were compared using identical GP priors, hyperpriors, and HMC-NUTS sampling procedures.

\begin{figure}
    \centering
    \includegraphics{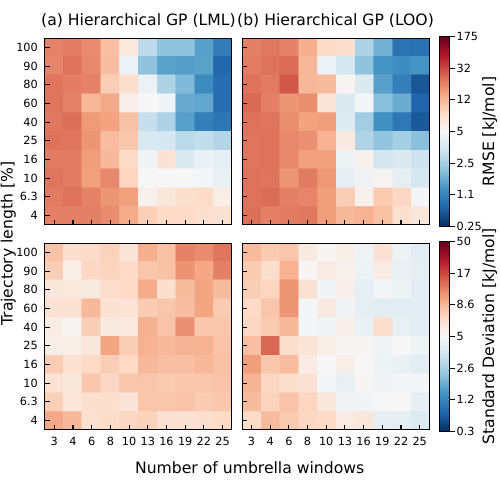}
    \caption{Mean RMSE and SD ablation grid heat maps comparing marginal likelihood and LOO pseudo-likelihood generalized hyperposterior constructions.}
    \label{fig:mlvsloo}
\end{figure}

A comparison between the marginal likelihood and LOO hyperparameter objectives at different data conditions is shown in Figure~\ref{fig:mlvsloo}.
The reconstructions obtained from the two objectives are similar for the posterior mean. 
The primary distinction between the two approaches emerged in their uncertainty estimates. 
Hyperposteriors constructed using the LOO pseudo-likelihood produced uncertainty estimates that more consistently tracked the observed accuracy of the reconstructed free energy surfaces across the data-ablation study.
On the other hand, the marginal likelihood objective actually exhibits an increase in predictive uncertainty as more data are obtained.

This behavior is expected from the underlying objectives. 
The marginal likelihood rewards hyperparameters that provide a globally self-consistent explanation of the observed data under the GP model. 
In contrast, the LOO pseudo-likelihood directly evaluates predictive performance by repeatedly assessing how well the model predicts observations omitted during training. 
Because UQ is fundamentally a predictive task, the LOO objective naturally places greater emphasis on predictive reliability than on explaining the training observations themselves.

These results suggest that the choice of hyperposterior objective function can influence the practical usefulness of uncertainty estimates even when its effect on the reconstructed mean is small. 
Consequently, for predictive UQ we recommend the LOO pseudo-likelihood as the default objective for hyperposterior inference in GP-based free energy reconstruction.

\section{Risks of Hyperposterior Optimization}\label{si:hyperopt}

In many GP applications, hyperparameters are optimized using marginal likelihood or a predictive objective (such as LOO), with uninformative (flat) hyperpriors.
Optimization locates the \textit{maximum a posteriori} (MAP),
\begin{equation}
\hat{\boldsymbol{\theta}}_{\mathrm{MAP}}=\arg\max_{\boldsymbol{\theta}}
p(\boldsymbol{\theta}\mid\boldsymbol{y}),
\end{equation}
or, in the case of the LOO objective,
\begin{equation}
\hat{\boldsymbol{\theta}}_{\mathrm{LOO}}=\arg\max_{\boldsymbol{\theta}}
p_{\mathrm{LOO}}(\boldsymbol{y}\mid\boldsymbol{\theta})
p(\boldsymbol{\theta}).
\end{equation}

\begin{figure}[h]
    \centering
    \includegraphics{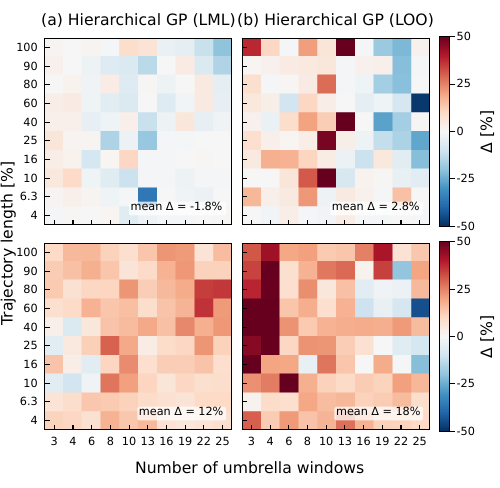}
    \caption{Percentage difference ($\Delta$) in mean RMSE and standard deviation between propagated hyperposteriors and MAP estimation. Blue indicates a decrease in the plotted quantity and red indicates an increase.}
    \label{si:mapvssample}
\end{figure}

Here we compare GP-based free energy reconstructions obtained using optimized hyperparameters to HMC-NUTS hyperposterior sampling, see Figure \ref{si:mapvssample}. 
The mean inferred by the two approaches was similar.
However, notable differences emerged in the uncertainty estimates. 
Hyperposterior sampling consistently produced broader predictive distributions, particularly in regions of sparse sampling where multiple hyperparameter configurations remained compatible with the available data.

From a practical perspective, the choice between optimization and hyperposterior sampling therefore depends on the intended application. 
When the primary goal is reconstruction of the mean, optimization is a reasonable and computationally efficient approximation.
However, UQ is non-negligibly impacted by neglecting uncertainties in the underlying hyperparameters, particularly in low-data regimes. 

We note that, for the systems considered in this work, the additional computational cost associated with HMC-NUTS remained negligible compared to the cost of generating MD trajectories.
The principal challenge is therefore not the computational expense of the Bayesian inference itself, but rather the scaling of hyperparamter-marginalized GP models to increasingly flexible kernel representations and higher-dimensional CV spaces.

\section{Metadynamics}\label{si:metad}

\subsection{Molecular Simulation Details}

A bilayer patch containing 128 1,2-dioleoyl-phosphatidylcholine (DOPC) molecules was generated using CHARMM-GUI Membrane Builder \cite{jo_charmm-gui_2008, wu_charmm-gui_2014, jo_charmm-gui_2009}. 
The system was solvated and the ions were added to neutralize the system and mimic 0.15M NaCl solution. 

All simulations were performed using GROMACS 2026.1 \cite{abraham_gromacs_2015, pall_tackling_2015, van_der_spoel_gromacs_2005}. 
The membrane was equilibrated using CHARMM36 \cite{huang_charmm36_2013} force field and the GROMACS input parameters recommended by CHARMM-GUI \cite{brooks_charmm_2009, lee_charmm-gui_2016}. 
To the equilibrated system, one phenol molecule was added and the system was equilibrated for another 20\,ns. 
The parameters of the phenol molecule were obtained using CHARMM Small Molecule Library \cite{kim_charmm-gui_2017}. 

The extended Lagrangian metadynamics simulation was performed using a PLUMED 2.9.3 plugin \cite{bonomi_promoting_2019, tribello_plumed_2014}. 
As a CV, we employed the distance between the center of mass (COM) of the membrane and the COM of the phenol molecule ranging from -1 to 5\,nm. 
The phenol was restrained using a flat-bottom potential with the force constant of 1000\,kJ/mol/nm$^2$. 
The temperature was set to 303.15\,K. 
The coupling constant of the extended Lagrangian was 10.000\,kJ/mol/nm$^2$ and the relaxation time was 1\,ps. 
The Gaussian potentials were applied with a rate of 0.004\,kJ/mol per picosecond and their widths were set to 0.1\,nm. 
The length of the metadynamics simulation was 1\,$\upmu$s.

In all simulations, the stochastic velocity rescaling thermostat \cite{bussi_canonical_2007} was used with a reference temperature of 303.15 K. 
The pressure of 1 atm was maintained by the stochastic cell rescaling barostat \cite{bernetti_pressure_2020}. 
The Verlet cutoff scheme was used \cite{verlet_computer_1967}. 
Force-switch cutoff of the van der Waals interactions was used between 1.0 and 1.2 nm. 
Long-range electrostatics was computed with the particle mesh Ewald method \cite{darden_particle_1993} with the cutoff of explicit short-range electrostatic interactions at 1.2 nm. 
An integration step of 2 fs was used in all simulations, with the bonds to hydrogen atoms converted to rigid constraints using the LINCS algorithm \cite{hess_lincs_1997} and water molecules constrained using SETTLE \cite{miyamoto_settle_1992}. 
Periodic boundary conditions were applied in all directions.

\subsection{Metadynamics Binning Procedure}

Unlike umbrella sampling, metadynamics trajectories are not naturally partitioned into independent sampling windows and therefore require \textit{post hoc} discretization of the history-dependent bias into histogram and derivative observations. 
The choice of discretization is not unique and can influence both the GP reconstruction and the inferred hyperparameter distributions. 
If the discretization is too fine, each observation contains too few data points, leading to unreliable estimates of the histogram densities and mean forces. 
This manifests as inflated inferred noise hyperparameters, particularly $\sigma_\mathrm{d}$. 
Conversely, if the discretization is too coarse, many data points are merged into a single observation, reducing the spatial resolution of the observations and producing unnecessarily smooth reconstructions with increased predictive uncertainty. 
The objective is therefore to employ the finest discretization that remains statistically well supported by the available data.

The histogram and derivative discretizations are controlled by the number of histogram bins, $n_\mathrm{h}$, and derivative windows, $n_\mathrm{d}$, respectively. 
For the histogram observations, increasing $n_\mathrm{h}$ improves spatial resolution because the stiff coupling between the auxiliary coordinate $z$ and the physical collective variable $\xi$ confines the sampled $\xi$ values to a narrow region within each bin. 
However, increasing $n_h$ also enlarges the GP covariance matrix, eventually making GP inference computationally prohibitive. For this study, values above approximately $n_\mathrm{h}=200$ became impractical computationally, while values below $n_\mathrm{h}\approx100$ noticeably degraded the predictive mean and uncertainty. 
Within the interval $100\leq n_\mathrm{h}\leq200$, the reconstruction was largely insensitive to the precise choice of $n_\mathrm{h}$.

The derivative discretization is instead limited primarily by the available simulation data. 
Choosing too few windows ($n_\mathrm{d}\lesssim30$) smooths the mean-force observations and degrades the reconstruction, whereas choosing too many windows produces noisy estimates of the average coupling force because each window contains too few samples. 
This behavior is reflected by increasing values of the inferred derivative noise hyperparameter, $\sigma_\mathrm{d}$, during hyperparameter optimization. 
To avoid overconfident or unstable reconstructions, we selected discretizations for which the inferred histogram and derivative noise hyperparameters remained of comparable magnitude over the full range of trajectory lengths considered.

Based on these considerations, the interval of interest (0--4.5\,nm) was discretized into $n_\mathrm{h}=120$ histogram bins and $n_\mathrm{d}=40$ derivative windows. 
Each bin and window contained approximately the same number of simulation samples, differing only in width. 
This equal-occupancy discretization was chosen to promote approximately homoskedastic observation noise by maintaining similar statistical precision across observations.

\bibliography{references}

\end{document}